\documentclass[prb,preprintnumbers,amsfonts,amssymb,amsmath,floats,twocolumn,aps]{revtex4}

\usepackage[pdftex]{graphicx}
\usepackage[pdftex,colorlinks=true,linkcolor=blue,citecolor=blue,filecolor=blue,baseurl=empty]{hyperref}
\usepackage{dcolumn}
\usepackage{bm}
\usepackage{color}
\usepackage{amsmath}
\usepackage[utf8]{inputenc}

\newcommand{\be}{\begin{equation}}
\newcommand{\ee}{\end{equation}}
\newcommand{\lr}[1]{\left(#1\right)}
\newcommand{\matrixel}[3]{\big< #1 \vphantom{#2#3} \big| #2 \big| #3 \vphantom{#1#2} \big>} 
\newcommand{\ket}[1]{\big| #1 \big>} 
\newcommand{\avg}[1]{\big< #1 \big>} 
\newcommand{\abs}[1]{\left|#1\right|}
\newcommand{\norm}[1]{\left|#1\right|^2}
\newcommand{\rec}[1]{\frac{1}{#1}}
\newcommand{\pd}[2]{\frac{\partial #1}{\partial #2}} 
\newcommand{\anticommut}[2]{\{#1,#2\}}
\newcommand{\vt}{\lr{t}}
\newcommand{\veps}{\lr{\epsilon}}
\newcommand{\vepsprime}{\lr{\epsilon'}}
\newcommand{\aC}[1]{a^{\dagger}_{#1}}
\newcommand{\aA}[1]{a_{#1}}

\newcommand{\fA}[1]{f_{#1}}
\newcommand{\tmax}[0]{t_{\text{max}}}

\begin{document}

\title{Pump-probe Auger-electron spectroscopy of Mott insulators}

\author{Roman Rausch}
\affiliation{Department of Physics, University of Hamburg, Jungiusstra{\ss}e 9, 20355 Hamburg, Germany}
\affiliation{Department of Physics, Kyoto University, Kyoto 606-8502, Japan}

\author{Michael Potthoff}
\affiliation{Department of Physics, University of Hamburg, Jungiusstra{\ss}e 9, 20355 Hamburg, Germany}

\begin{abstract}
In high-resolution core-valence-valence (CVV) Auger electron spectroscopy from the surface of a solid at thermal equilibrium, the main correlation satellite, visible in the case of strong valence-electron correlations, corresponds to a bound state of the two holes in the final state of the CVV Auger process. 
We discuss the physical significance of this satellite in nonequilibrium pump-probe Auger spectroscopy by numerical analysis of a single-band Hubbard-type model system including core states and a continuum of high-energy scattering states.
It turns out that the spectrum of the photo-doped system, due to the increased double occupancy, shares features with the equilibrium spectrum at higher fillings.
The pumping of doublons can be watched when working with overlapping pulses at short $\Delta t$. 
For larger pump-probe delays $\Delta t$ and on the typical femtosecond time scale for electronic relaxation processes, spectra are hardly $\Delta t$-dependent, reflecting the high stability of bound two-hole states for strong Hubbard-$U$.
We argue that taking into account the spatial expansion of single-particle orbitals when these are doubly occupied, as described by the dynamical Hubbard model, produces an oscillation of the barycenter of the satellite as a function of $\Delta t$.
Pump-probe Auger-electron spectroscopy is thus highly sensitive to dynamical screening of the Coulomb interaction.
\end{abstract}

\maketitle

\section{Introduction}

High-resolution core-valence-valence (CVV) Auger-electron spectroscopy (AES) \cite{Aug25,Lan53,Pow88} can be employed as a tool to examine the electronic structure of solid surfaces. 
It complements photoemission spectroscopy \cite{FFW78,CL78,CH84,DHS03} as the standard technique to access the occupied  states close to the Fermi energy. 
Compared to photoemission, AES has the advantage of being directly sensitive to local electron correlations: \cite{VCM01}
Due to the core state involved in the transition, the CVV Auger process is predominantly local, i.e., the two final-state holes are mainly created at the same lattice site.
For a ``weakly correlated'' simple metal, the Coulomb interaction among the valence electrons can be accounted for on a mean-field level. 
In this case the final-state holes quickly delocalize and propagate essentially independently through the lattice. 
Disregarding matrix-element effects, \cite{Pot01b,PWN+00} the raw Auger spectrum is given by the occupied part of the self-convolution of the local density of the states. \cite{Lan53}
The spectrum from a ``strongly correlated'' metal, on the other hand, may look substantially different. 
A sufficiently strong Hubbard-type local interaction $U$ causes a two-hole bound state which moves as a compound object through the lattice \cite{RP17} and is visible is the Auger spectrum as a correlation satellite. \cite{RP16}
The important difference with respect to photoemission spectroscopy is that, for $U$ much larger than the valence bandwidth $W$, the satellite nearly takes the whole spectral weight.

A nice example is Ni, where the two-hole bound state, the famous ``6-eV satellite'', is not easily seen in photoemission \cite{GBP+77} while the Ni CVV Auger spectrum exhibits a strong correlation satellite. \cite{MJ80,Nol90,WPN00}
The Cini-Sawatzky theory \cite{Cin77,Saw77} developed in the late 1970s for systems with a completely filled valence band predicts the position of the satellite as compared to the center of gravity of the two-hole scattering continuum to be approximately given by $U$. 
This has been exploited in the past in an attempt to measure $U$ for transition metals and compounds. \cite{dBHS84,BMN+93} 

For the general case of a partially filled band, the theoretical description is challenging, even for simple models, such as the single-band Hubbard model. \cite{Gut63,Hub63,Kan63} 
Here, the Auger spectrum must be obtained from the local elements of the two-particle Green's function, and this task requires to deal with the typical obstacles of a full many-body problem in the strong-coupling case.
Various ideas have been used to generalize the Cini-Sawatzky theory, \cite{Cin79,GDDS81,DK84,OTSJ86,CV86,Nol90} including perturbative, exact-diagonalization, decoupling and diagrammatic re-summation techniques.
A numerically exact solution for the entire filling range of the one-dimensional Hubbard model could be provided recently. \cite{RP16}

In recent years, the idea of repulsively bound pairs of particles has also attracted the attention of researchers in the field of ultra-cold quantum-gases. \cite{WTL+06} 
As a consequence of energy and momentum conservation, there is no phase space available for a dissociation of a pair of bosons in a Bose-Hubbard model. \cite{Blo05,JZ05}
In case of a finite concentration of particles (bosons or fermions), a repulsively bound pair can only decay in a high-order scattering process involving several low-lying excitations of the continuum, and the lifetime is exponentially long in $U$, as verified experimentally \cite{SGJ+10} and theoretically. \cite{SGJ+10,LP13,CGK12}
On a short time scale, there is an initial  ``decay'', \cite{HP12,RP17} which is related to the energy-time uncertainty principle.

It is tempting to ask if the decay of a repulsively bound pair of final-state holes can be watched on the time axis by means of {\em time-resolved} CVV Auger spectroscopy.
While time-resolved photoemission \cite{IOM+03,PLL+06,WBC+11,BRP+15} has become popular in the last decade to study the time evolution of electronic excitations of materials characterized by strong electron correlations well beyond the linear-response regime, there has not much attention been paid to Auger-electron spectroscopy -- despite its high sensitivity to correlation effects.
Atomic systems represent a notable exception. \cite{DHK+02}

This is the starting point and motivation of the present paper. 
Here, we like to ask what be learnt from time-dependent AES from solid surfaces.
To this end we propose an exemplary theoretical description of time-dependent AES covering essential features of strongly-correlated systems and focus on the time-dependence of the Auger correlation satellite in particular.
To be specific, the single-band Hubbard model at half-filling and, as a variant, the dynamic Hubbard model \cite{Hirsch_2001,Hirsch_2002,Werner_Eckstein_2016} will be considered. 
At strong Hubbard interaction $U$, these represent generic models for a correlation-induced Mott insulator.
The present study profits from our previous work \cite{RP16} on the {\em equilibrium} two-hole spectrum of the Hubbard model at and off half-filling, where a comprehensive analysis of the various spectral features was provided. 

\begin{figure}[t]
\includegraphics[width=0.8\columnwidth]{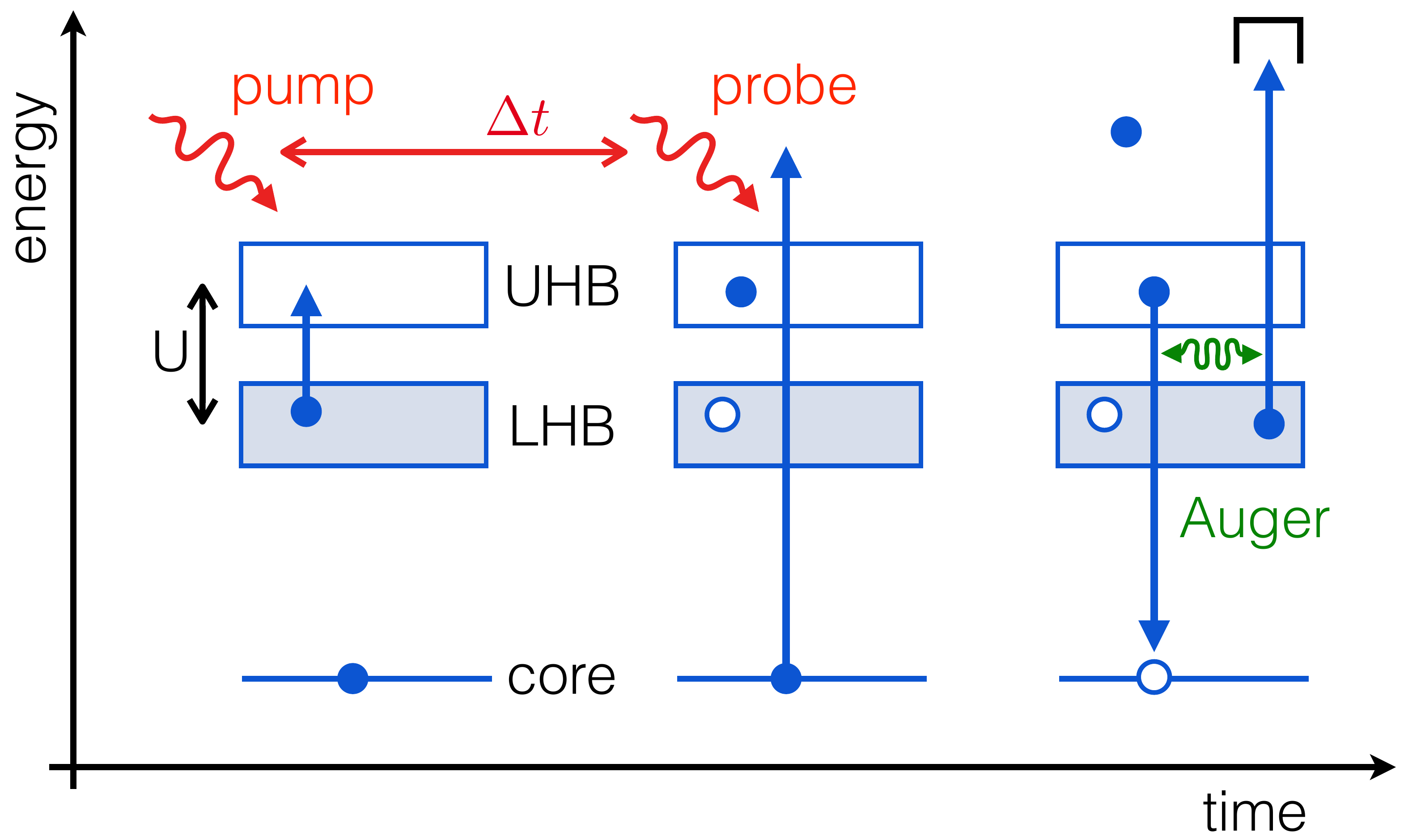}
\caption{
Sketch of pump-probe Auger electron spectroscopy from a Mott insulator with lower (LHB) and upper Hubbard band (UHB) separated by the Hubbard interaction $U$.
An optical pump pulse drives the system to a nonequilibrium state with an enhanced fraction of double occupancies.
The system evolves in time. 
After time $\Delta t$ a core hole is created by an X-ray probe pulse. 
The spectrum of the subsequent Auger decay is characteristic for the local correlated electronic structure at time $\Delta t$.
}
\label{fig:aes}
\end{figure}

The theory for the {\em nonequilibrium} case addresses a realistic experimental pump-probe setup, see Fig.\ \ref{fig:aes}.
We consider an optical pump pulse with a mean frequency of the order of a few eV to bridge the Mott gap and with a duration in the femtosecond regime. The pump drives the system out of equilibrium and generates an initial nonthermal conduction-electron state with an enhanced fraction of doubly occupied sites. 
This state evolves in time.
A relaxation of the electronic degrees of freedom generically takes place on the fs scale.
After a time delay $\Delta t$, a hole in an inner core shell is created by an X-ray probe pulse which opens a channel for the subsequent CVV Auger decay. 
The dependence of the Auger spectrum on the time delay $\Delta t$ reflects the time evolution of the conduction-electron state. 
Our main focus is on the intense Auger correlation satellite and particularly on its time-dependent position and weight.
Which kind of information on the electronic structure is encoded here?
This fundamental question can be answered within simplified Hubbard-type models for the correlated electronic structure of a Mott insulator augmented by core states and a continuum of high-energy scattering states using exact-diagonalization techniques.

In the following section \ref{sec:mod}, we introduce the model and the typical parameters involved in the calculations. 
Section \ref{sec:theory} outlines our theoretical approach. 
We first present results on the core-hole decay in Sec.\ \ref{sec:core} before discussing the pump-probe Auger spectra for the conventional Hubbard model in Sec.\ \ref{sec:pp} and the dynamic Hubbard model in Sec.\ \ref{sec:dyn}.
Our findings are summarized in Sec.\ \ref{sec:sum}.

\section{Model}
\label{sec:mod}

To be definite and to facilitate the numerical evaluation, the theory will be formulated for the one-dimensional single-band Hubbard model at half-filling as a prototypical model for the correlated electronic structure of a Mott insulator. \cite{Geb97}
An extension to higher dimensions and multi-orbital models is straightforward in principle.
The Hamiltonian is given by 
\be
H_{\rm Hub}(t) = H_{\rm hop}(t) + U \sum_{i} n_{i\uparrow} n_{i\downarrow} \: ,
\ee
where $n_{i\sigma} = c^{\dagger}_{i\sigma} c_{i\sigma}$ is the occupation-number operator, and $c^{\dagger}_{i\sigma}$ creates an electron at lattice site $i = 1, ..., L$ with spin projection $\sigma=\uparrow, \downarrow$. 
The strength of the local Hubbard interaction is given by $U$.
The tight-binding hopping term of the Hamiltonian $H_{\rm hop}(t)$ with the nearest-neighbor hopping-matrix element $T$ reads
\be
H_{\text{\text{hop}}} (t) 
= 
-T \sum_{i\sigma} \lr{e^{iA_{\text{pmp}}\lr{t}} c^{\dagger}_{i\sigma} c_{i+1\sigma} + \text{h.c.}} \: .
\ee
The energy scale (and with $\hbar \equiv 1$ the time scale as well) is fixed by setting $T \equiv 1$.
The hopping Hamiltonian is time dependent as it includes the coupling of the light field to the electronic system, i.e., the pump pulse. 
This coupling is described in a standard manner in the velocity gauge by using the Peierls substitution.
Thereby, the hopping matrix element between the valence orbitals acquires a complex phase which is given by the time-dependent vector potential $A_{\text{pmp}}(t)$. 

In addition, we consider the core degrees of freedom involved in the Auger transition, i.e.,
\be
H_{\rm core} = -E_c \sum_{i\sigma} n^{\rm (f)}_{i\sigma} + U_{cc} \sum_{i} n^{\rm (f)}_{i\uparrow} n^{\rm (f)}_{i\downarrow} + U_{cv} \sum_{i\sigma\sigma'} n^{\rm (f)}_{i\sigma} n_{i\sigma'} \: ,
\ee
which describes fully localized core states with bare one-particle energy $-E_c$, the core-core Coulomb interaction $U_{cc}$ and the core-valence Coulomb interaction $U_{cv}$, assumed as local for simplicity.
Accordingly, $n^{\rm (f)}_{i\sigma} = f^{\dagger}_{i\sigma} f_{i\sigma}$ is the occupation-number operator for the core states, and $f^{\dagger}_{i\sigma}$ creates a core electron at lattice site $i$ with spin projection $\sigma$.

For both the pump pulse and the probe pulse (index $\gamma = \text{pmp, prb}$), the electric field with frequency $\Omega_{\gamma}$ and amplitude $\mathcal{E}_{\gamma}^{(0)}$,
\be
\mathcal{E}_{\gamma}(t) = \mathcal{E}_{\gamma}^{(0)} \cos\big[\Omega_{\gamma} \lr{t-t_{\gamma}}\big] s_{\gamma}(t) \: ,
\label{eq:E_t}
\ee
is assumed to have a Gaussian shape with an envelope function,
\be
s_{\gamma}(t) = \exp\left[-\rec{2}\lr{\frac{t-t_{\gamma}}{\tau_{\gamma}}}^2\right] \: ,
\ee
characterized by $t_{\gamma}$ and $\tau_{\gamma}$.
The corresponding vector potential is obtained by integration:
\be
A_{\gamma}(t) = - \int_{-\infty}^t dt' \, \mathcal{E}_{\gamma}\lr{t'} \: .
\label{eq:A_t}
\ee
Note that for a harmonic pulse with Gaussian envelope, as assumed here, this results in the complex error function which can be numerically efficiently evaluated using the Faddeeva package. \cite{FaddeevaPackage}
Furthermore, we define the pump-probe delay as the time difference between the respective pulse peaks: $\Delta t = t_{\text{prb}} - t_{\text{pmp}}$.

The continuum of high-energy scattering states above the vacuum level is represented by
\be
H_{\rm scatt} = \int d\epsilon~ \rho\lr{\epsilon} \epsilon~ a^{\dagger}_{\sigma}\lr{\epsilon} a_{\sigma}\lr{\epsilon} \: .
\ee
Here, $\rho\lr{\epsilon}$ is the corresponding density of states.
$a^{\dagger}_{\sigma}\lr{\epsilon}$ creates an electron with spin projection $\sigma$ and energy $\epsilon$ in the scattering continuum, and $a_{\sigma}\lr{\epsilon}$ is the corresponding annihilator.
These operators obey fermionic anticommutation relations in the form $\anticommut{\aA{\sigma}\veps}{\aC{\sigma'}\vepsprime} = \rho\veps^{-1} \delta_{\sigma\sigma'} \delta\lr{\epsilon-\epsilon'}$ and are thus dimensionless.
Note that we assume two different continua, one for the scattering states occupied by the emitted photoelectron and another one for the Auger electron with one-particle energies $\epsilon_{\phi}$ and $\epsilon\equiv\epsilon_A$, respectively (usually suppressed for better readability). 
We assume that the two continua are well-separated in energy and that the photoelectron and the Auger electron can be treated as distinguishable particles. 
Effects that result from their non-distinguishability, such as post-collision interactions, \cite{Schuette_etal_2012} are beyond the scope of this work.

The system is probed by core-level photoemission, and the corresponding term in the Hamiltonian describing the probe pulse is given by 
\be
V_{\text{prb}}\lr{t} = \mathcal{E}_{\text{prb}}\vt \sum\limits_{i\sigma} \int d\epsilon~ \rho\lr{\epsilon} d_i\veps a^{\dagger}_{\sigma}\lr{\epsilon}\fA{i\sigma} \: ,
\label{eq:vx}
\ee
Here, we employ the length gauge and assume a constant dipole matrix element $d_i\veps \approx d_0$.
The subsequent Auger transition is mediated by a Coulomb-interaction term between core, valence and scattering states:
\begin{equation}
H_{A} = \sum_{i\sigma} \int d\epsilon~ \rho\veps \lr{ U^A_{i}\veps a^{\dagger}_{\sigma}\veps f^{\dagger}_{i-\sigma} c_{i-\sigma} c_{i\sigma} + h.c. } \; , 
\label{eq:va}
\end{equation}
where again the transition matrix element is approximated by a constant, i.e., $U^A_{i} \veps \approx U_A$.
The full Hamiltonian then reads:
\begin{equation}
H_{\text{el}}(t) = H_{\text{Hub}}\vt + H_{\rm core} + H_{\text{scatt}} + H_A + V_{\text{prb}}(t) \; .
\label{eq:ham}
\end{equation}

\begin{widetext}

\section{Theory}
\label{sec:theory}

Assuming that the probe is weak and using low-order time-dependent perturbation theory in $V_{\text{prb}}\vt$, one straightforwardly arrives at the following one-step spectroscopic formula: \cite{Rausch_2017,Gunnarsson_Schoenhammer_1980}
\be
I_{\text{AES}}\lr{\epsilon} = \:  
2\pi\rho_{\phi} \lim_{\tmax\to\infty} \int\limits_0^{\tmax} dt' \:s^2_{\text{prb}} (t')
\cos^2 [\Omega_{\rm prb} (t'-t_{\rm prb}) ] \;
\matrixel{\Psi_{\phi} (\tmax,t')}{n(\epsilon)}{\Psi_{\phi}(\tmax,t')} \; ,
\label{eq:spec}
\ee
\end{widetext}
where $\rho\lr{\epsilon_{\phi}} \approx \rho_{\phi} = \text{const}$ is the (constant) density of states of the photoelectron continuum, $n\lr{\epsilon} = \sum_{\sigma} \aC{\sigma}\veps \aA{\sigma}\veps$ is the occupation-number operator for the Auger-electron continuum, 
and where 
\be
  \ket{\Psi_{\phi} (\tmax,t')} = U(\tmax,t') V^{\lr{1}}_{\text{prb}} \, U (t',0) \ket{\Psi(0)} \: ,
\label{eq:psi}
\ee
is a state which is probed at time $t'$ (note that $t'$ is integrated over in Eq.\ (\ref{eq:spec})). 
Furthermore, $U(\cdot,\cdot)$ is the time-evolution operator corresponding to the Hamiltonian (\ref{eq:ham}) including the pump pulse, but with $V_{\text{prb}}(t) \equiv 0$.
At time $t=0$, the system is in its ground state $\ket{\Psi(0)}$.
Obviously, without a pump ($\mathcal{E}_{\text{pmp}}^{(0)}=0$), the first time evolution in Eq.\ (\ref{eq:psi}) becomes trivial, and one would just create a core hole in the ground state at time $t'$. 
For small $U_A$ this would essentially recover the Cini-Sawatzky theory of Auger emission, \cite{Cin77,Saw77} with the sole difference of having a probe pulse rather than a continuous beam.

In Eq.\ (\ref{eq:psi}) the perturbation describing the probe pulse reduces to
\be
V_{\text{prb}}^{(1)} \equiv d_0 \sum\limits_{i\sigma} \fA{i\sigma} \: 
\label{eq:v1}
\ee
when making use of the sudden approximation, i.e., assuming that $a_{\sigma}(\epsilon) \ket{\Psi(0)} = 0$.
For an X-ray probe, calculations can be further simplified by exploiting that $\Omega_{\text{prb}}$ is very large on the energy scale of the conduction band set by the hopping $T$. 
Replacing the fast oscillating cosine square term stemming from Eq.\ (\ref{eq:E_t}) in the integral Eq.\ (\ref{eq:spec}) by its time-average, 
\be
\cos^2\big[\Omega_{\text{prb}}\lr{t-t_{\text{prb}}}\big] \to \overline{\cos^2}\big[\Omega_{\text{prb}}\lr{t-t_{\text{prb}}}\big] = \frac{1}{2} \; ,
\label{eq:sin_avg}
\ee
only the envelope of the probe pulse remains.
Since this a positive-definite function, it can be used as a weight function to create custom Gauss-Legendre integration weights using the Golub-Welsh algorithm, \cite{Golub_Welsch_1969} so that only few integration points $N_t$ are necessary:
\be
\int dt' s^2_{\text{prb}}(t') f(t') \approx \sum_{i=1}^{N_t} w_i f(t_i) \; .
\ee
This has also the advantage that the pulse can be made arbitrarily short without numerical problems. 
In practice, having just $N_t=3$ integration points turns out as sufficient.

Another useful approximation relies on the locality of the photoemission and Auger process, i.e., the creation of the core hole and the subsequent Auger decay basically take place at the same lattice site (say, at $i=i_0$) opposed to a picture with a coherent superposition over all lattice sites. 
This approximation has also been justified {\em a posteriori} by a general calculation which shows that extra-atomic interferences cancel out when the inverse photoelectron momentum $k_{\phi}$ is much larger than the lattice spacing $a$, i.e., if $k_{\phi}^{-1} \gg a$. \cite{Abraham-Ibrahim_etal_1978, Rausch_2017} 
Furthermore, because of spin degeneracy, we can limit ourselves to just one spin direction (say, $\sigma = \uparrow$), and hence we may replace:
\be
V_{\text{prb}}\lr{t} \to \mathcal{E}_{\text{prb}}\vt d_{i_0}\veps \int\rho\lr{\epsilon}d\epsilon~ a^{\dagger}_{\uparrow}\lr{\epsilon}\fA{i_0\uparrow}
\ee
and
\be
H_{A} \to U_A \int d\epsilon~ \rho\veps \lr{ a^{\dagger}_{\downarrow}\veps f^{\dagger}_{i_0,\uparrow} c_{i_0\uparrow} c_{i_0\downarrow} + h.c. }
\ee
in Eqs.\ (\ref{eq:vx}) and (\ref{eq:va}), respectively, and similarly in Eq.\ (\ref{eq:v1}).
Hence, the equilibrium spectral function, for instance, is of the type $\avg{c^{\dagger}_{i_0\downarrow}c^{\dagger}_{i_0\uparrow} \delta\lr{\epsilon-H} c_{i_0\uparrow}c_{i_0\downarrow}}$ rather than $\sum_{ij}\avg{c^{\dagger}_{i\downarrow}c^{\dagger}_{i\uparrow} \delta\lr{\epsilon-H} c_{j\uparrow}c_{j\downarrow}}$. 

The Auger-electron scattering states could be eliminated in the case of weak $U_A \ll T$ in very much the same way as for the photoelectron scattering states, which lead us to Eq.\ (\ref{eq:v1}), namely truncating the Taylor expansion of the time evolution operator after the first order, the $a^{(\dagger)}_{\sigma}\lr{\epsilon}$ disappear by commutation. 
We note, however, that the resulting equations are not necessarily easier to handle, since they involve several time propagations and subsequent integrations. 
It is therefore in fact easier to account for the Auger-electron continuum explicitly, which can be achieved without too much more computational effort and also allows for computations that are non-perturbative with respect to $U_{A}$. 
In practice, we adapt the mapping to a semi-infinite chain using orthogonal polynomials, as introduced by Chin et al. \cite{Chin_etal_2010} 
The procedure is described in detail in the Appendix \ref{app:OrthoPolyMap}.

\section{Core-hole lifetime and maximal propagation time}
\label{sec:core}

To test the above framework and to determine some of the parameters, we study the finite core-hole lifetime due to the Auger decay. 
The discussion of the complete pump-probe setup is shifted to the next section, and thus, for the time being, we assume the pump to be switched off. 
Using exact diagonalization, we calculate the total cross-section $\sigma = \int d \epsilon ~ \rho \veps I_{\text{AES}}(\epsilon)$ [see Eqs.\ (\ref{eq:spec}) and (\ref{eq:sigma_gen})] as a function of $U_A$ with a finite probe pulse rather than a continuous beam. 
The density of states is assumed as constant, $\rho \veps = \rho_{A} = \mbox{const}$.

Assuming that the core-hole lifetime $\tau_{c}$ is long compared to the pulse duration, $\tau_c \gg \tau_{\text{prb}}$, there are two analytically accessible limits: 
(i) For the fully occupied band, $n \equiv \sum_{\sigma} \langle n_{i\sigma} \rangle = 2$, the time propagation can be carried out exactly for any $U_A$ and $U_{cv}$. 
(ii) For weak $U_A$ and $U_{cv}=0$, perturbation theory can be used. 

We take a finite cutoff time $\tmax$ [see Eq.\ (\ref{eq:spec})] and find the same result for both cases, namely:
\be
  \sigma\lr{\tmax} = 2\pi \rho_{\phi} A_s \big|d_0\big|^2 \avg{n_{i\uparrow}n_{i\downarrow}} \lr{1-e^{-2\Gamma \tmax}},
\label{eq:sigma_exp}
\ee
(see Refs.\ \onlinecite{Rausch_2017,Gunnarsson_Schoenhammer_1980} for details) or, for $\Gamma\tmax \ll 1$:
\be
\sigma\lr{\tmax} \approx 2\pi \rho_{\phi} A_s \big|d_0\big|^2 \avg{n_{i\uparrow}n_{i\downarrow}} \cdot 2\Gamma\tmax \: ,
\label{eq:sigma_lin}
\ee
with the width \cite{Gunnarsson_Schoenhammer_1980}
\be
2\Gamma = 2\pi \rho_A \big|U_A\big|^2 \; .
\ee
$A_s$ is the area under the probe envelope squared, which for the given Gaussian pulse is given by
\be
A_s = \frac{1}{2} \int dt'~ s^2\lr{t'} = \frac{\sqrt{\pi}}{2} \tau_{\text{prb}} \: .
\ee
The core-hole lifetime can be defined as
\be
\tau_c = \frac{1}{2\Gamma} \: .
\ee

\begin{figure}[t]
\includegraphics[width=0.95\columnwidth]{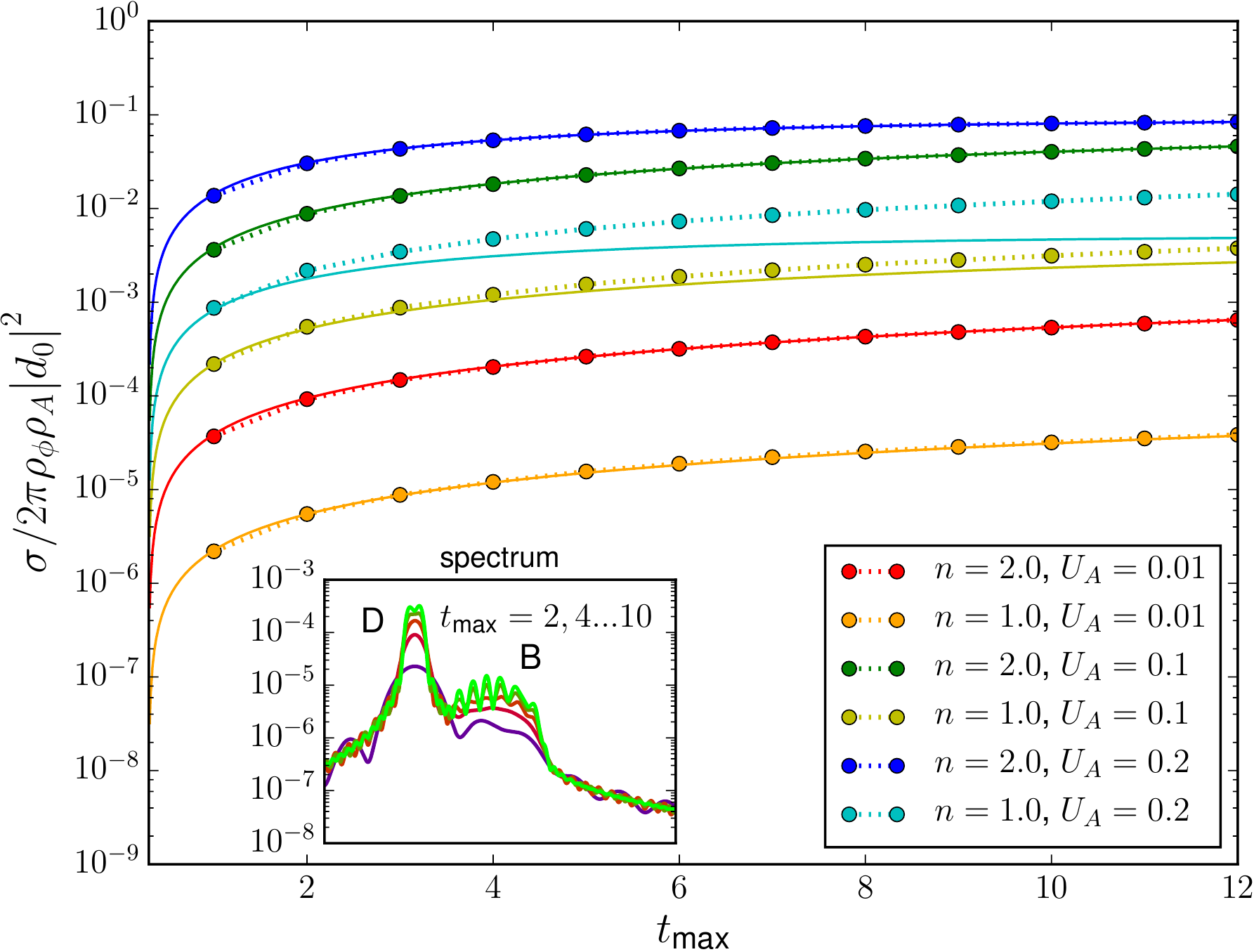}
\caption{
{\em Main figure:} Total AES cross-section as a function of the cutoff time $\tmax$ as obtained numerically (dots and dashed lines) compared with the analytical formula (solid lines) given by Eq.\ (\ref{eq:sigma_exp}), valid for the full band $n=2$ and all $U_{A}$ as well as for small $U_A$ and all band fillings $n$. 
Additional parameters: $L=10$, $U=6$, $U_{cv}=0$, $\tau_{\text{prb}}=0.1$, $L_{\text{ch}}=200$ (to describe the continumm, see Appendix \ref{app:OrthoPolyMap}). 
{\em Inset:} Convergence of the spectrum for $n=2$ and $U_A=0.1$ as a function of $\tmax$ and eventual appearance of finite-size effects (see green line for $\tmax=10$). 
``B'': band-like part. ``D'': doublon satellite (see text).
The energy (and with $\hbar \equiv 1$ also the time scale) is fixed by $T=1$.
}
\label{fig:sigma}
\end{figure}

We have checked the validity of the above formula by comparing with the fully numerical evaluation, see Fig.\ \ref{fig:sigma}. 
Some deviations are expected as the core hole starts to decay already {\em during} probe pulse, but these get smaller with shorter pulse duration $\tau_{\rm prb}$. 
For $\tau_{\rm prb} = 0.1$, the agreement is almost perfect in the two mentioned limits. 

Calculations are done for a finite cutoff time $\tmax$.
Note that the cross-section simply grows in the cutoff time (linearly at first, not directly visible on the log scale). 
The convergence of the Auger spectrum with increasing $\tmax$ is seen in the inset of Fig.\ \ref{fig:sigma}. 
Obviously, finer spectral features are resolved at larger $\tmax$. 
In fact, using a finite $\tmax$ even before the spectrum is converged is not necessarily unphysical, but simply smoothens out the result further and can be regarded as finite numerical resolution. 
A system of $L=10$ sites (in a linear chain geometry with periodic boundary conditions) is already enough to observe the basic spectral features, i.e., the band-like part (``B'') and the doublon satellite (``D'') at $U=6$; see also the discussion in the following section and Ref.\ \onlinecite{RP16}.

To get the core-hole lifetime for the non-trivial case of $n=1$, $U_{cv}\neq 0$ and large or moderate $U_A$ to be studied here, we calculate the spectrum for several $\tmax$ and employ a two-parameter fit assuming the same functional dependence (and shifting the time-zero to the probe peak):
\be
\sigma\lr{\tmax} \sim  \alpha \lr{1-e^{-2\Gamma \lr{t_{\text{max}}-t_{\text{prb}}}}} \: .
\label{eq:fit_exp}
\ee
Since in the linear regime ($U_A \lesssim 0.01$ and $\tmax - t_{\rm prb} \lesssim 10$), an exponential fit is ill-defined, we rather use a linear one-parameter fit:
\be
\sigma\lr{\tmax} \sim  2\Gamma \lr{t_{\text{max}}-t_{\text{prb}}} \; .
\label{eq:fit_lin}
\ee

\begin{figure}[t]
\includegraphics[width=0.95\columnwidth]{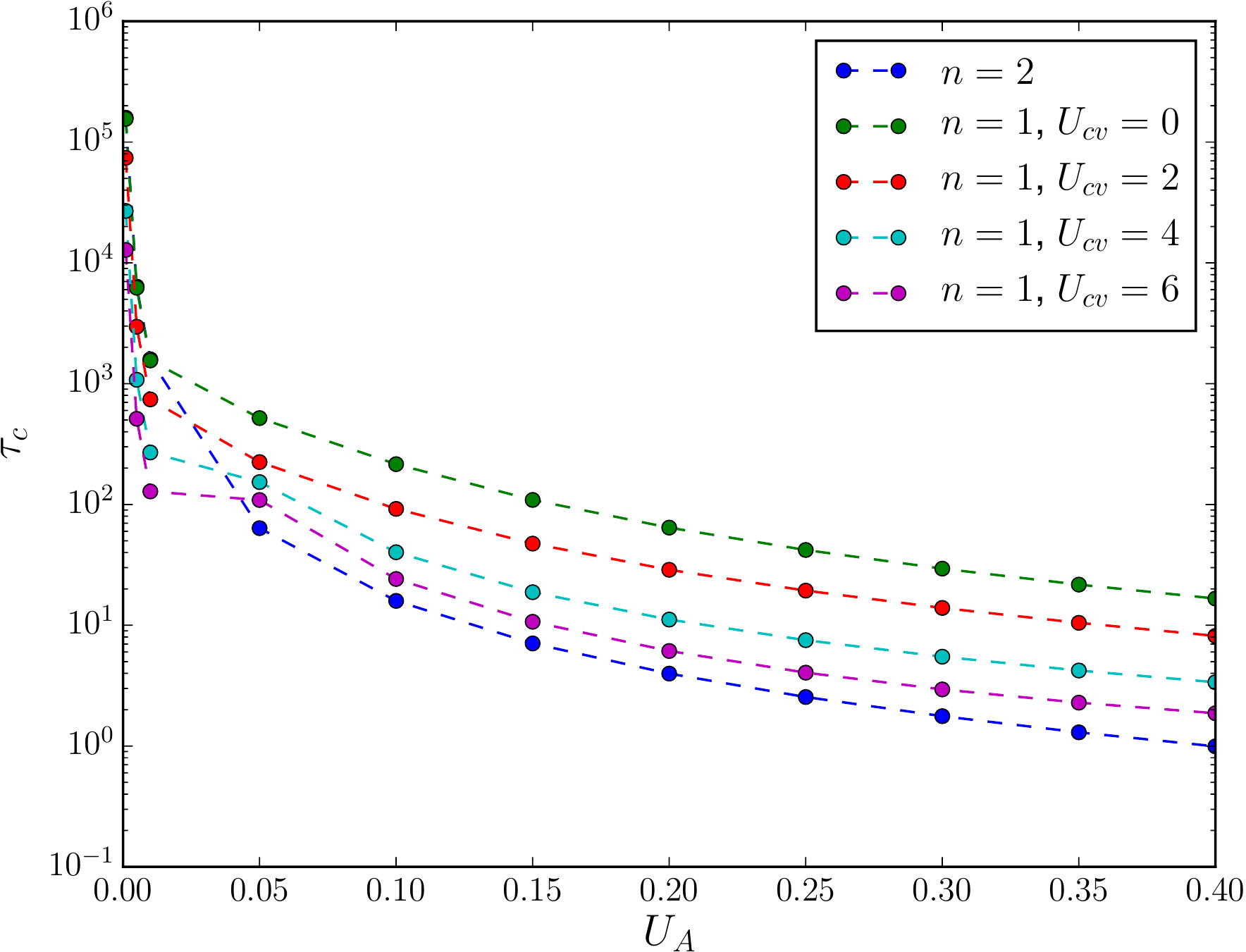}
\caption{
Core-hole lifetime $\tau_c=1/2\Gamma$ as a function of $U_{A}$ extracted from the exponential fit Eq.\ (\ref{eq:fit_exp}) for $U_A > 0.01$ and from the linear fit Eq.\ (\ref{eq:fit_lin}) for $U_A < 0.01$. 
The fits are done in the range $\tmax = 1,2, \ldots 12$ (see Fig.\ \ref{fig:sigma}). Note that the lifetime for $n=2$ does not depend on $U_{cv}$. Other parameters as in Fig.\ \ref{fig:sigma}.
}
\label{fig:tau}
\end{figure}

The results are shown in Fig.\ \ref{fig:tau}. 
Since there are less conduction electrons available to fill the core hole in the half-filled case $n=1$, its lifetime is at least an order of magnitude higher as compared to $n=2$. 
However, increasing $U_{cv}$ reduces the lifetime, as the core-hole potential is quickly screened by surrounding electrons.
As a function of $U_A$ (for $U_A \gtrsim 0.01$), we find a power-law behavior of the lifetime, $\tau_{c} \propto U_{A}^{\alpha}$, where the exponent increases from $\alpha \approx 1.6$ for $U_{cv} = 0$ to $\alpha = 2$ for strong $U_{cv}$, i.e., for strong core-hole screening we recover the $n=2$ result.

\section{Pump-probe spectra for the pure Hubbard model}
\label{sec:pp}

We proceed to the calculation of pump-probe spectra and at first discuss the pump-pulse parameters in Eq.\ (\ref{eq:E_t}). 
The pump frequency and the intensity of the pulse will be chosen as to maximize the resulting double occupancy, see below.
The pump duration is fixed at $\tau_{\text{pmp}}=1$, i.e., at the time scale set by the inverse nearest-neighbor hopping $1/T$. 
A typical value of $T \sim 100$~meV would correspond to about $\tau_{\rm pmp} \sim 40$~fs.

\begin{figure}[t]
\includegraphics[width=0.95\columnwidth]{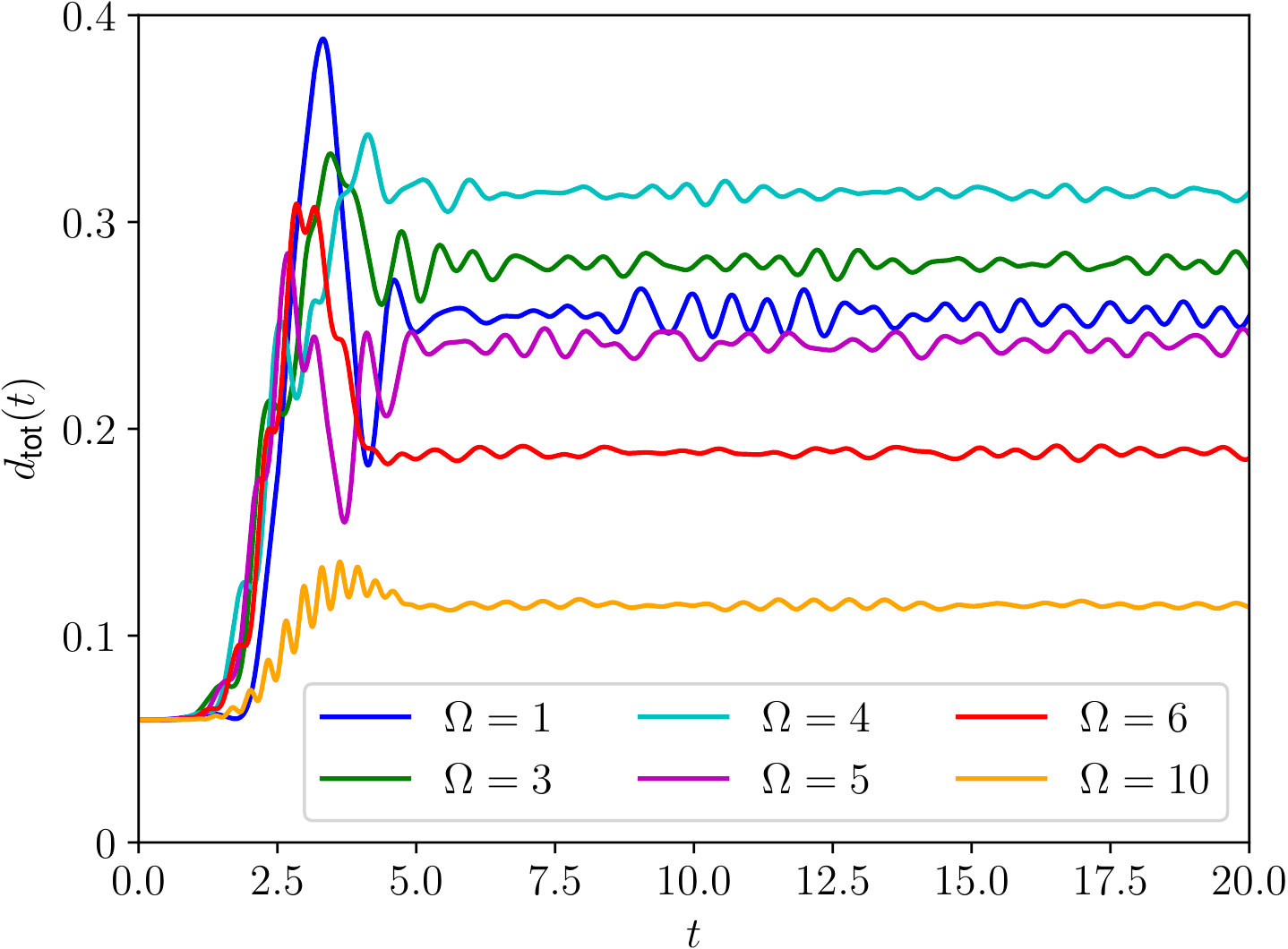}
\caption{
Total double occupancy after a pump [see Eq.\ (\ref{eq:E_t})] of duration $\tau_{\text{pmp}}=1$ for various values of $\Omega_{\text{pmp}}$ as indicated. 
Further parameters: $L=10$, $U=6$, $\mathcal{E}^{(0)}_{\text{pmp}} = 8$, periodic boundary conditions. 
}
\label{fig:optimalPulse}
\end{figure}

For the calculations, i.e., for the time-propagation during the pump pulse in particular, see Eq.\ (\ref{eq:psi}), we use the CFET algorithm \cite{Alvermann_2011, Alvermann_Fehske_Littlewood_2012} based on the Magnus series expansion. \cite{Magnus_1954}
Results for the double occupancy 
\be
d_{\text{tot}}\vt = \frac{1}{L} \sum_i \matrixel{\Psi\vt}{n_{i\uparrow}n_{i\downarrow}}{\Psi\vt}
\label{eq:dtot}
\ee
are shown in Fig.\ \ref{fig:optimalPulse}.

The optimal $\Omega_{\text{pmp}}$ is expected to be given by the energy of an excitation from the middle of lower Hubbard band to the middle of upper Hubbard band. 
For $U=6$ and half-filling this amounts to $\Omega_{\text{pmp}} \approx W/2 + \Delta + W/2 \approx 5.45$ where $W=4$ is the noninteracting band width and where the charge gap $\Delta \approx 1.446$ at $U=6$ is taken from the Bethe ansatz solution. \cite{Ovc69}
Based on the data for $d_{\text{tot}}\vt$ for frequencies $\Omega_{\text{pmp}}$ in the vicinity of this value, it turns out that the best choice is achieved with a slightly smaller value, $\Omega_{\text{pmp}} \approx 4$, see Fig.\ \ref{fig:optimalPulse}. 
With higher frequencies, the resulting double occupancy goes down again, since the electronic system becomes unable to react to such fast driving.
The driving field strength $\mathcal{E}^{(0)}_{\text{pmp}}$ has been varied simultaneously and maximizes the double occupancy for $\mathcal{E}^{(0)}_{\text{pmp}} \approx 8$. 
These optimal values could be somewhat affected by finite-size effects.

\begin{figure}[t]
\includegraphics[width=0.95\columnwidth]{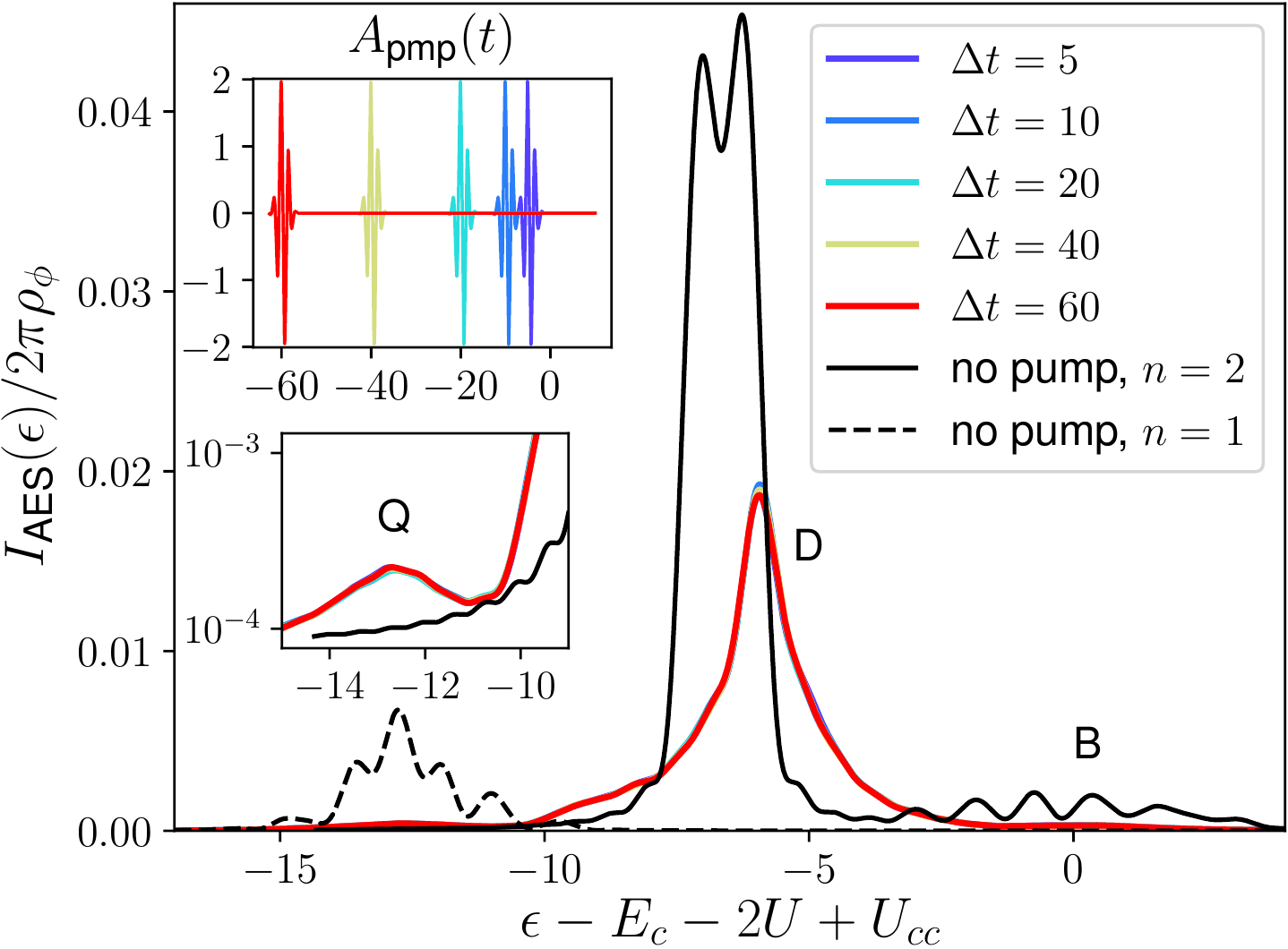}
\caption{
Pump-probe Auger spectra of the half-filled Hubbard model for various delay times $\Delta t$ at $U=6$. 
Calculations for $L=10$, periodic boundary conditions. 
Comparison with the equilibrium case for the full band $n=2$ (solid line) and the half-filled band $n=1$ (dashed line) for the same parameters. 
The upper inset shows the vector potential of the pump $A_{\text{pmp}}\vt$.
The probe starts at $t=0$, i.e., $t_{\rm prb}=0$.
The lower inset is a magnification of the quadruplon peak at small kinetic energies (``Q''). 
``D'' denotes the doublon satellite, ``B'' is the band-like part, see Ref.\ \onlinecite{RP16} and discussion in the  text.
}
\label{fig:AESpp_k=0}
\end{figure}

Fig.\ \ref{fig:AESpp_k=0} shows the pump-probe Auger spectrum for several delay times $\Delta t$ at $U=6$ and $n=1$, compared with the equilibrium case, i.e., for the pump pulse switched off. 
We first discuss the equilibrium case for $n=2$ (solid black line).
The broad peak at highest kinetic energies of the Auger electron is the band-like part (``B''), where the doublon dissociates into two final-state hole which propagate more or less independently and delocalize.
Note that there are some spurious oscillations due to the finite system size.
The maximal kinetic energy that the Auger electron can obtain is given by the difference of $E_0\lr{N-1}$, the ground-state energy with one core hole, and $E_0\lr{N-2}$, the ground-state energy with two holes in the valence band. 
For the full band and strong $U$, this is given by $\epsilon_{\text{max}} = 2U+W+E_c-U_{cc}$, whereby $\epsilon=2U+E_c-U_{cc}$ marks the barycenter of the band-like part which has a width of $2W$.
Separated by an energy of about $U$, the spectrum features a satellite corresponding to the case that the doublon does not decay and forms a repulsively bound object (``D'').
At fillings $n < 2$ (not shown), an additional quadruplon satellite (``Q'') with very low spectral weight appears, separated yet again by about $U$ from the doublon one. \cite{RP16} 
This corresponds to a bound state of four holes in the system. 
At half-filling $n = 1$, all these features have collapsed into a single peak, see the dashed line for $n=1$, where again some spurious finite-size effects are visible.

We are now prepared to make a comparison with the time-resolved (pump-probe) case. 
Here, the spectroscopy is performed on a photo-doped system at half-filling. 
Due to the increased double occupancy, the spectrum shares certain features with the equilibrium spectrum at higher fillings. 
In particular, we observe an intense peak at the energy position of the doublon satellite. 
Its flat upper shoulder reaches into the band-like part ``B'', while the lower shoulder exhibits a peak at the quadruplon energy (``Q''), albeit an order or magnitude smaller, which will be difficult to observe experimentally.
The doublon satellite ``D'', on the other hand, is the dominating spectral structure.

However, with regard to the delay time $\Delta t$, we note that there is almost no dependency whatsoever; the spectra calculated for $\Delta t = 5$ up to $\Delta t= 60$ basically fall on top of each other. 
The physical reason is that the doublons created during the pump pulse are extremely stable at $U=6$.
Their decay time is expected to increase exponentially with increasing Coulomb interaction $U$ in the strong-coupling regime. \cite{SGJ+10, Eckstein_Werner_2011} 
Thus, the double occupancy, to which the Auger process is sensitive, more or less stays constant after the pump is over (see also the right inset of Fig.\ \ref{fig:optimalPulse}), so that it does not matter when, on the typical (femtosecond) timescale of electron dynamics, the probe is applied.

This result has important implications:
Apparently, the Auger spectrum is highly robust even when recorded from a system in an excited nonequilibrium state, its overall shape being simply determined by the value of the double occupancy $d_{\text{tot}}$ and its position by the value of $U$. 
Hence, pump-probe spectra will only yield time-resolved information if either of these quantities fluctuates in time. 

The former, however, is a quasi-conserved quantity in the interesting strong-coupling regime, except, of course, {\em during} the pump pulse.
Hence, there is the possibility to work with overlapping pulses, i.e., probe the system while pumping of doublons is still active.
Therewith, the $\Delta t$-dependence would reflect the time dependence of the doublon-pumping process. 

As an even more interesting alternative, one may look into cases with a time-dependent (effective) interaction $U(t)$.
A time-dependent Hubbard-$U$ can be realized in a number of ways, in particular within the dynamic Hubbard model. 
This possibility is discussed in the following.

\section{Pump-probe spectra for the dynamic Hubbard model}
\label{sec:dyn}

The dynamic Hubbard model \cite{Hirsch_2001, Hirsch_2002, Werner_Eckstein_2016} was introduced by Hirsch to account for the effect that whenever two electrons are on the same site, they will not simply occupy the same single-electron orbital. 
Rather, the orbital will relax and expand due to the Coulomb interaction between them. 
To account for this mechanism on the level of the Hubbard model, one can add a bosonic degree of freedom that couples to the double occupancy. 
Here we adopt the simple approach where this is given by a classical harmonic oscillator. 
Hence, our extended model now reads as:
\be
H(t) = H_{\text{el}}(t) + H_{\text{bos}} + H_{\text{int}} \: .
\ee
Here, $H_{\text{el}}(t)$ comprises the electronic system, Eq.\ (\ref{eq:ham}), discussed before. 
Furthermore, 
\be
H_{\text{bos}} = \frac{\omega_0^2}{2} \sum_i p_i^2 + \rec{2} \sum_i q_i^2
\ee
is the Hamilton function of the classical oscillators with frequency $\omega_{0}$, and
\be
H_{\text{int}} = k \sum_i q_{i} O_{i}
\label{eq:Hint}
\ee
is the interaction term with coupling strength $k$. 
We use rescaled coordinates and momenta as described in the Appendix \ref{app:DynHubProp}.

The choice of the operator $O_i$ leads to different physical situations: 
For $O_i = n_i$ we obtain the semiclassical Hubbard-Holstein model, which can be of interest for a study of doublon decay due to the presence of phonons, while for $O_i = n_{i\uparrow} n_{i\downarrow}$ we obtain the dynamic Hubbard model. 
This is characterized by a modified, {\rm effective} Coulomb interaction, 
\be
U_{i,\text{eff}} (t) = U + k q_i (t) \: ,
\label{eq:ueff}
\ee
which in general is time-dependent.

\begin{figure}[t]
\includegraphics[width=0.95\columnwidth]{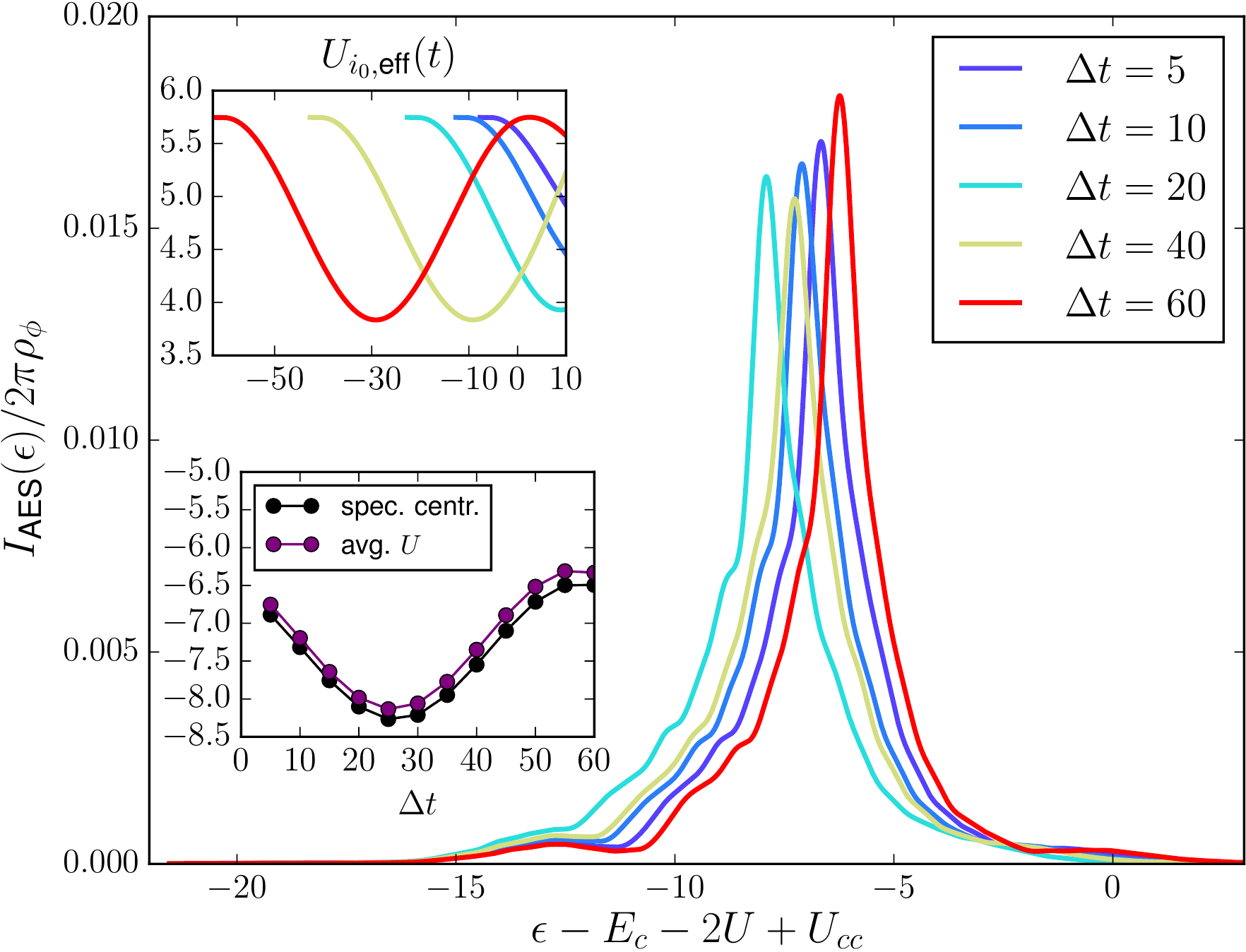}
\caption{
Pump-probe Auger spectra of the half-filled dynamic Hubbard model for various delay times $\Delta t$ at $U=6$. 
Calculations for the same parameters as in Fig.\ \ref{fig:AESpp_k=0}, for coupling strength $k=2$ and for $\omega_0=0.1$ (adiabatic case).
{\em Upper inset:} 
The effective Coulomb interaction as a function of time, see Eq.\ (\ref{eq:ueff}). 
Note: $\tau_{\rm prb}=0$.
{\em Lower inset:} 
The barycenter of the spectrum, see Eq.\ (\ref{eq:spec.centr}), compared with the time-averaged averaged $U_{i,\text{eff}}\vt$, see Eq.\ (\ref{eq:avgU}).
}
\label{fig:AESpp_adiabatic}
\end{figure}

\begin{figure}[b]
\includegraphics[width=0.95\columnwidth]{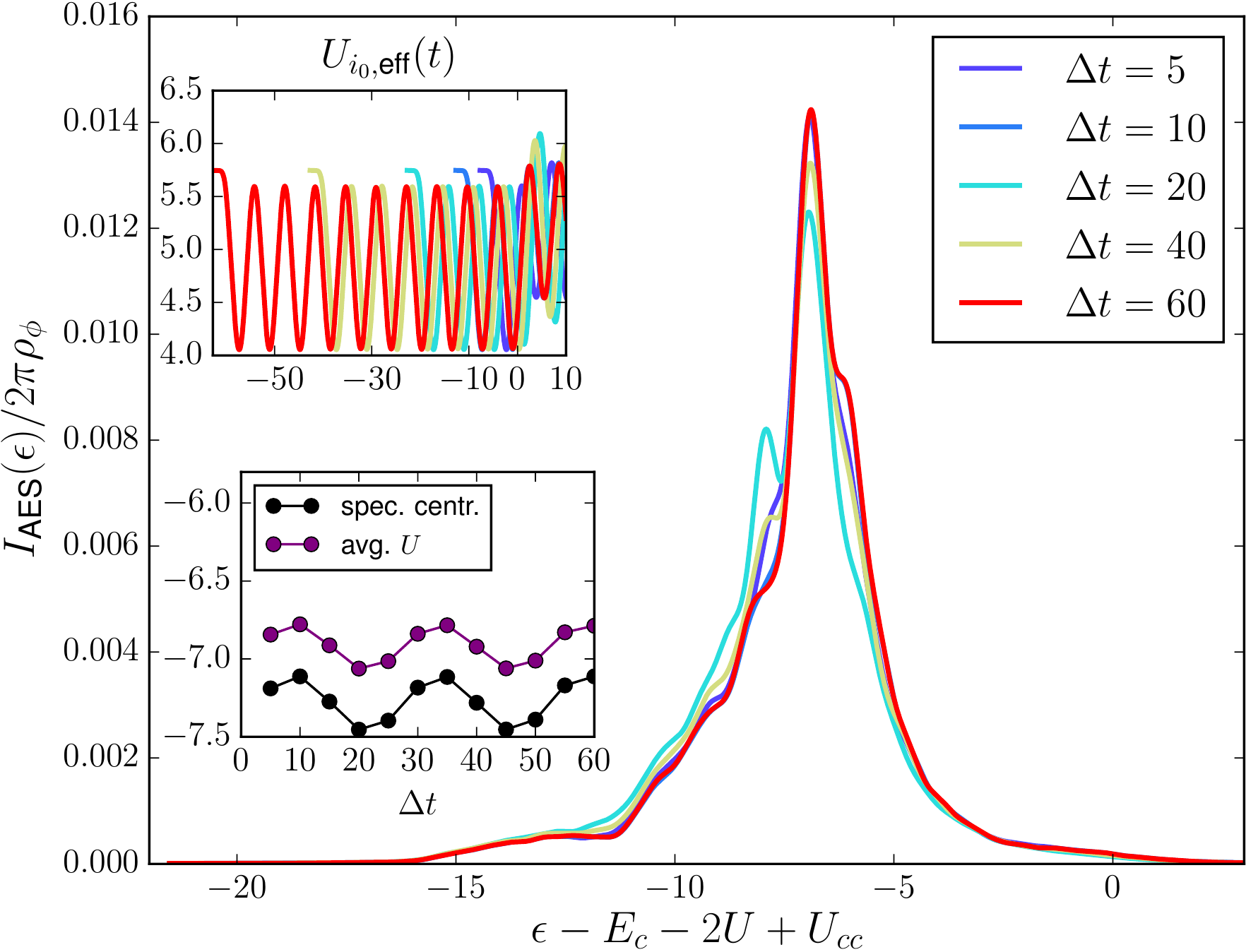}
\caption{
The same as Fig.\ \ref{fig:AESpp_adiabatic}, but for the antiadiabatic case with $\omega_0=1$.
}
\label{fig:AESpp_antiadiabatic}
\end{figure}

To time-propagate the quantum-classical hybrid system, we exactly map the quantum degrees of freedom onto an additional ``virtual'' set of coupled classical harmonic oscillators. 
The dynamics of this large nonlinear classical system is then treated using the standard splitting integrator approach.\cite{Ruth_1983,McLachlan_1995,Grochowski_Lesyng_2003, BCM08, Lubich_2008} 
The detailed procedure is outlined in Appendix \ref{app:DynHubProp}.

In Fig.\ \ref{fig:AESpp_adiabatic} we show the Auger spectra of the dynamic Hubbard model for the adiabatic case. 
This can be realized with a frequency $\omega_0=0.1$ much smaller than the bare electronic time scale which is given by the inverse nearest-neighbor hopping $1/T=1$. 
Calculations are done for a moderately strong coupling $k=2$. 
With this choice of the coupling constant the system is deep in the homogeneous phase as can be seen from the equilibrium phase diagram of the model discussed in the Appendix \ref{app:DynHubPD}. 
During the pump pulse ($\tau_{\text{pmp}}=1$), electrons are excited into the upper Hubbard band, i.e., double occupations are created. 
The doubly occupied orbitals then start to oscillate, basically with the frequency $\omega_0$, after some initial reaction to the pulse.
As shown in the upper inset of the figure, this means that there is relatively long timescale $\tau_{\text{osc}} = 2\pi/\omega_0 \approx 62.8$, on which the effective interaction [see Eq.\ (\ref{eq:ueff})] varies.

It now matters at which point of this cycle the Auger decay is initiated via the probe pulse.
Namely the resulting reduction or enhancement of $U_{i,\text{eff}}\vt = U + k q_i\lr{t}$ is responsible for a corresponding energy shift of the correlation satellite of the Auger spectrum to lower or higher binding energies, respectively, see the main part of Fig.\ \ref{fig:AESpp_adiabatic}.
This observation can be made even quantitative.
At $U=6$ and for strong $U$ in general, the correlation satellite basically takes almost the whole spectral weight, so that its position is to a very good approximation given by the barycenter of the complete Auger spectrum:
\be
\overline{I}_{\text{AES}} = \int d\epsilon~ \epsilon~ I_{\text{AES}}\lr{\epsilon} \Big/ \int d\epsilon~ I_{\text{AES}}\lr{\epsilon} \: .
\label{eq:spec.centr}
\ee
Since the core hole takes a certain time to decay, we compare the barycenter with $U_{i,\text{eff}}\vt$, averaged over some time interval $\rm t_{\rm eff}$,
\be
  \overline{U}_{i,\text{eff}} = \frac{1}{t_{\text{eff}}} \int_0^{t_{\text{eff}}} dt~ U_{i,\text{eff}} (t) \: ,
  \label{eq:avgU}
\ee
see the lower inset of Fig.\ \ref{fig:AESpp_adiabatic}.
Taking $t_{\text{eff}}=10$ appears to produce the best results, so that the barycenter even matches $\overline{U}_{i,\text{eff}}$ almost perfectly. 

Finally, in Fig.\ \ref{fig:AESpp_antiadiabatic} we show results for the antiadiabatic case, with $\omega_0=1$. 
The effective interaction now oscillates on the much shorter timescale of $\tau_{\text{osc}} \approx 6.28$ (upper inset). 
This oscillation is in fact fast enough to complete at least one cycle within the lifetime of the core hole.
Consequently, $\overline{U}_{i,\text{eff}}$ almost averages out to $U_{i,\text{eff}}\lr{0}$, with a weak remaining oscillation. 
Again, the barycenter of the spectrum follows $\overline{U}_{i,\text{eff}}$ almost perfectly (see lower inset of Fig.\ \ref{fig:AESpp_antiadiabatic}).

We conclude that the frequency $\omega_{0}$ manifests itself directly in the delay-time dependence of the main correlation satellite in pump-probe Auger spectroscopy.
A significant effect that should be observable is expected if the period is much longer than the core-hole lifetime, i.e., if $\tau_{\text{osc}}  = 2\pi / \omega_{0} \gg \tau_{c}$.

\section{Discussion and summary}
\label{sec:sum}

To summarize, let us compare CVV Auger-electron spectroscopy from an equilibrium electronic state with nonequilibrium pump-probe AES. 
We consider a strongly correlated system with a valence electronic structure, described by a Hubbard model with interaction $U$ larger than the bare bandwidth $W$.
In the equilibrium case, the local sensitivity of AES can be employed to determine the Hubbard-$U$ as the energy distance of the correlation satellite from the band-like part of the spectrum, since the satellite is shifted in energy by the two-hole binding energy of about $U$. 
The satellite almost takes the whole spectral weight if $U\gg W$ which makes AES much more sensitive to correlations as compared to single-electron photoemission spectroscopy.
This argument builds on the simple single-band Hubbard model and must be refined for real materials (i) in case of systems with partially filled bands where an explicit many-body calculation is necessary to determine the precise satellite position and (ii) by taking into account all relevant low-energy electronic degrees around the Fermi energy, i.e., in by considering a multi-orbital Hubbard-type model.

For the nonequilibrium case, we have seen that the Auger process starts from a photo-doped nonequilibrium state with enhanced double occupancy. 
In a mode with overlapping pump and probe pulses, the system is probed while still pumping doublons, and the dependence on the pump-probe delay time $\Delta t$ would reflect the time dependence of the doublon-pumping process. 
After the pump, and in the interesting strong-coupling regime $U\gg W$, however, the double occupancy is a quasi-conserved quantity due to missing phase space for decay in a single scattering event. 
The important consequence is that the pump-probe Auger spectrum is basically independent of $\Delta t$.
Actually, one would expect at least a weak $\Delta t$ dependence due to a relaxation of doublons to the bottom of the upper Hubbard band. \cite{Eckstein_Werner_2011,Werner_Eckstein_2016}
To see this effect, however, one would have to consider much larger systems or use mean-field-type methods working directly in the thermodynamical limit.

Interestingly, the nonequilibrium spectrum exhibits features that are characteristic of the equilibrium spectrum from a system with higher band filling and thus higher equilibrium double occupancy. 
This has been verified for a Mott insulator at half-filling where the calculation shows a quadruplon peak in the nonequilibrium spectrum at much higher binding energies compared to the doublon satellite, which is impossible in the equilibrium case at half-filling but shows up at higher fillings. \cite{RP16}
As the location of the main doublon satellite and additional features, such as a quadruplon peak, sensitively depend on the average number of doubly occupied sites, pump-probe AES could be used as a tool to measure the doublon concentration in the initial transient state as a function of the pump parameters. 

It is well known \cite{WC16} that, in the context of quantitative studies for real materials, the single-band Hubbard model is, at best, an effective low-energy theory whose parameters must be determined in a downfolding process by integrating out higher-energy degrees of freedom, eventually by starting from an all-electron Hamiltonian in the continuum.
This downfolding generates a dynamically screened Coulomb interaction, i.e., in the simplest case a frequency- or time-dependent Hubbard-$U$, 
which leads to important observable effects such as a reduction of the effective bandwidth and of the spectral weight of low-energy excitations as has been pointed out recently. \cite{CWV+12}
A first step to capture the effects of dynamical screening can already be done with the dynamic Hubbard model \cite{Hirsch_2001} involving a single bosonic mode of frequency $\omega_{0}$ as has been considered here.
The screening frequency $\omega_{0}$ is the essential ingredient of the effective low-energy theory as it determines the observable bandwidth renormalizations. \cite{CWV+12}
We have demonstrated that pump-probe AES is in fact directly sensitive to the dynamical screening process, subject to the condition that $2\pi / \omega_{0}$ is larger than the core-hole lifetime.
The essential observation is that the position of the intense correlation satellite, as a function of the pump-probe delay $\Delta t$, just oscillates with the screening frequency $\omega_{0}$. 
This is easily understood in a simple picture, where the Hubbard interaction, after integrating out the bosonic mode, becomes time-dependent and oscillates with frequency $\omega_{0}$, such that the satellite position $\sim U_{\rm eff}(t)$ must oscillate with basically the same frequency.
Simple estimates as well as state-of-the-art many-body calculations \cite{Hirsch_2001,CWV+12} show that $2\pi / \omega_{0}$ is in the relevant femtosecond regime for strongly correlated materials, such as Mott insulators, and is thus well accessible to modern pump-probe setups.

\acknowledgments

We would like to thank M.\ Martins and W.\ Wurth (University of Hamburg) for instructive discussions.
Financial support of this work through the Deutsche Forschungsgemeinschaft within the Sonderforschungsbereich 925 ``Light induced dynamics and control of correlated quantum systems'' (project B5) is gratefully acknowledged. 

\appendix

\section{Orthogonal polynomial mapping}
\label{app:OrthoPolyMap}

A set of orthogonal polynomials $\{ p_{n}\}$ is defined either via a recurrence relation or via a positive-definite weight function, see Ref.\ \onlinecite{Chin_etal_2010}.
We can treat the density of states of a continuum $\rho\veps$ (having a compact support) as such a weight function and define the scalar product between two polynomials $p_n$ in the following fashion:
\be
\left<p_n,p_m\right> = \norm{U_A} \int d\epsilon~ \rho\veps p_n\veps p_m\veps = \delta_{nm} \: .
\ee
One can then define new discrete creation and annihilation operators in the following way:
\be
b^{\dagger}_{n\sigma} = U_A \int d\epsilon~ \rho\veps p_n\veps \aC{_\sigma}\veps \: .
\ee
Note that they are still fermionic: $\anticommut{b_{n\sigma}}{b^{\dagger}_{n'\sigma'}} = \delta_{nn'}\delta_{\sigma\sigma'}$. 
Using the orthogonality of the polynomials, the inverse transformation is found to be
\be
\aC{\sigma}\veps = U_A \sum_{n=0}^{\infty} p_n\veps b^{\dagger}_{n\sigma} \: .
\ee
The scattering continuum now becomes a semi-infinite chain,
\begin{eqnarray}
H_{\text{sc}} &=& \sum_{n=0}^{\infty}\sum_{\sigma} \alpha_n b^{\dagger}_{n\sigma}b_{n\sigma} 
\nonumber \\
&+& \sum_{n=0}^{\infty}\sum_{\sigma} \sqrt{\beta_{n+1}} \lr{ b^{\dagger}_{n\sigma}b_{n+1,\sigma} + \mbox{H.c.} } \; ,
\end{eqnarray}
while the Auger term couples all sites to the first one of the chain:
\begin{equation}
H_A = \abs{\pi_0} \sum_{i\sigma} \lr{b^{\dagger}_{0\sigma}f^{\dagger}_{i,-\sigma} c_{i,-\sigma} c_{i\sigma} + \mbox{H.c.}} \: .
\end{equation}
Here, $\alpha_n$ and $\beta_n$ are the recurrence coefficients of the corresponding monic polynomials $\pi_n$, and $\abs{\pi_0} = \sqrt{\left<\pi_0,\pi_0\right>} = \big|U_A\big| \sqrt{\int\rho\veps d\epsilon}$ is the norm of the first one. 
In the case of a constant density of states, $\rho\veps=\text{const}=\rho_{A}$, and a support limited between $\epsilon_{\text{min}}$ and $\epsilon_{\text{max}}$, the polynomials $p_n$ are given as shifted Legendre polynomials whose recurrence relations are explicitly given by
\be
\begin{split}
\alpha_n &= b \: , \\
\beta_n  &= \frac{a^2}{4-n^{-2}} \: , \quad n\geq1 \; ,
\end{split}
\ee
with $a = \lr{\epsilon_{\text{max}}-\epsilon_{\text{min}}}/2$ and $b = \lr{\epsilon_{\text{max}} + \epsilon_{\text{min}}}/2$, so that $\abs{\pi_0} =  U_A \sqrt{2a\rho_{A}}$.
We also note that the total cross-section $\sigma$ can be written as
\be
\begin{split}
\sigma &= \int d\epsilon~ \rho\veps I_{\text{AES}}\lr{\epsilon} \\
       &= 2\pi\rho_{\phi} \int\limits_0^\infty dt' \:s^2_{\text{prb}}\lr{t'}\: \matrixel{\Psi_{\phi}\lr{t'}}{N_{\text{ch}}}{\Psi_{\phi}\lr{t'}} \: ,
\end{split}
\label{eq:sigma_gen}
\ee
where
\be
N_{\text{ch}} = \sum_{n\sigma} b^{\dagger}_{n\sigma} b_{n\sigma}
\ee
is the total particle number operator within the chain.

For the given problem, the initial state is a product state of the system ground state and the chain vacuum. 
Hence, cutting the chain to a finite length $L_{\text{ch}}$ does not introduce any error. 
For the time evolution, however, one has to ensure that the occupation of the last site $\sum_{\sigma} \avg{b^{\dagger}_{L_{\text{ch}}-1,\sigma} b_{L_{\text{ch}}-1,\sigma}}$ remains small to have a negligible error. 
Otherwise a reflection from the open end will cause unphysical results.

Note the relation to energy-time uncertainty: 
With increasing time an increasing number of higher-order polynomials is necessary to describe the dynamics of particle traveling through the chain. 
This corresponds to a successively better resolution of spectral features.

It is important to realize that $\epsilon_{\text{max}}$ and $\epsilon_{\text{min}}$ do not have to encompass the full support of the spectrum. 
We can choose to concentrate on a small spectral window only and measure the spectrum there. 
This implies a smaller effective hopping amplitude, both into the chain $\abs{\pi_0}$ and along the chain $\beta_n$ (due to the smaller width $a$). Hence, one needs less chain sites and a smaller Hilbert space, but in turn, more spectral windows are needed to reconstruct the full spectrum. 
Because of the perturbative treatment of $V_{\text{prb}}$, there can only be a single electron occupying the scattering states, i.e., in the chain. 
Hence, the Hilbert-space dimension after the probe is given by $\dim\mathcal{H} = \dim\mathcal{H}^{\lr{N-1}}_{\text{el}} + L_{\text{ch}} \times \dim\mathcal{H}^{\lr{N-2}}_{\text{el}}$. 
In the extreme case we can actually have $L_{\text{ch}}=1$ (no increase of Hilbert space at all, apart from the sum), which only allows for a polynomial of degree zero (a constant) and quite a lot of spectral windows are needed to get the full spectrum. In practice, a compromise between memory and operations can be struck, setting $L_{\text{ch}}$ as high as possible for the available memory resources.

\section{Time propagation of the dynamic Hubbard model}
\label{app:DynHubProp}

We couple our quantum system to classical oscillators described by the Hamiltonian function
$H_{\rm bos} + H_{\rm int}$ with
\begin{eqnarray}
H_{\text{bos}} &=& \rec{2m} \sum_i \tilde{p}_i^2 + \frac{m\omega_0^2}{2} \sum_i \tilde{q}_i^2 \; ,
\nonumber \\
H_{\text{int}} &=& \lambda \sum_i \tilde{q}_i O_i \: .
\end{eqnarray}
Using the standard rescaling of the oscillator coordinates, i.e., $k = \lambda/\sqrt{m\omega_0^2}$, $q_i = \sqrt{m\omega_0^2} ~ \tilde{q}_i$, $\tilde{p}_i = \sqrt{m\omega_0^2} ~ p_i$, allows us to rewrite the terms as
\begin{eqnarray}
H_{\text{bos}} &=& \frac{\omega_0^2}{2} \sum_i p_i^2 + \rec{2} \sum_i q_i^2 \equiv T_{\rm bos}^{p} + V_{\rm bos}^{q}
\: , 
\nonumber \\
H_{\text{int}} &=& k \sum_i q_i O_i \equiv H_{\rm int}^{q} \: , 
\label{eq:hint}
\end{eqnarray}
where $O_i = n_{i\uparrow} n_{i\downarrow}$.
The superscripts $q,p$ indicate the respective dependencies.

Following Ref.\ \onlinecite{Grochowski_Lesyng_2003}, we choose an orthonormal basis $\{ \ket{m} \}$ of the many-body Hilbert space to expand the state vector in the Schr\"odinger picture, 
\be
\ket{\Psi(t)} = \sum_m c_m(t) \ket{m} \: ,  
\ee
which can be used to define a classical Hamiltonian function for all the occurring degrees of freedom $\left\{\vec{q},\vec{p},\vec{c}\right\}$ where $\vec{q} \equiv (q_{1}, ..., q_{L})^{T}$ and $\vec{p}$ analogously and $\vec{c} \equiv (c_{1}, c_{2}, ... c_{M})^{T}$, and $M$ is the Hilbert-space dimension:
\begin{eqnarray}
\mathcal{H} 
&=& H_{\text{bos}} + \langle \Psi | ( H_{\rm el} + H_{\rm int} ) | \Psi \rangle
\nonumber \\
&=& T^p_{\text{bos}} + V^q_{\text{bos}} + \vec{c}^{\dagger} ( T_{\rm el} + V^q_{\rm int} ) \vec{c}
\: .
\end{eqnarray}
Here $T_{\rm el}$ and $V^q_{\rm int}$ are the matrix representations of $H_{\rm el}$ and $H_{\rm int}$ with respect to the chosen basis.

With the Liouville-Operator, given by
\begin{eqnarray}
\mathcal{L} 
&=& \pd{\mathcal{H}}{\vec{p}} \pd{}{\vec{q}} - \pd{\mathcal{H}}{\vec{q}} \pd{}{\vec{p}} 
-i \pd{\mathcal{H}}{\vec{c}^*} \pd{}{\vec{c}} + \mbox{H.c.} \: ,
\end{eqnarray}
we get the equations of motion for the classical degrees of freedom, 
\begin{eqnarray}
\dot{q}_i & = & \pd{ \mathcal{H} }{p_i} = \omega_0^2p_i \: , 
\nonumber \\
\dot{p}_i & = & -\pd{ \mathcal{H} }{q_i} = - q_i - k\avg{n_{i\uparrow}n_{i\downarrow}}_t
\; ,
\label{eq:motion_pq}
\end{eqnarray}
as well as for the $c$ degrees of freedom, which reproduces the Schr\"odinger equation:
\be
\dot{\vec{c}} = \mathcal{L}_{\text{qm}} \vec{c} = -i \lr{T_{\text{el}} + V^q_{\text{int}}} \vec{c} \: .
\ee
The so-called PICKABACK splitting, \cite{Grochowski_Lesyng_2003} 
$\mathcal{L} = \mathcal{L}_{qT} + \mathcal{L}_{pV}$ with 
\begin{eqnarray}
\mathcal{L}_{qT} &=&  \pd{T^p_{\text{bos}}}{\vec{p}} \pd{}{\vec{q}} -i \pd{T_{\text{el}}}{\vec{c}^*} \pd{}{\vec{c}} + \mbox{H.c.} \: , 
\nonumber \\
\mathcal{L}_{pV} &=& -\pd{(V^q_{\text{bos}} 
+ 
\vec{c}^{\dagger} V^q_{\rm int} \vec{c}
)}{\vec{q}} \pd{}{\vec{p}} -i \pd{V^q_{\text{int}}}{\vec{c}^*} \pd{}{\vec{c}} + \mbox{H.c.} \: ,
\nonumber \\
\end{eqnarray}
allows us to perform alternate propagations with these two Liouville operators:
\begin{eqnarray}
\exp [(\mathcal{L}_{qT} + \mathcal{L}_{pV}) \delta t ] 
&=& \prod_{l=1}^r \exp\lr{\mathcal{L}_{pV}a_l\delta t} \exp\lr{\mathcal{L}_{qT}b_l\delta t} 
\nonumber \\
&+& O\lr{\lr{\delta t}^{r+1}}
\: .
\end{eqnarray}
Here, we employ the fourth-order method ($r=4$), characterized by the choice $a_1=a_4=1/\lr{2x}$, $a_2=a_3=\lr{x-1}/\lr{2x}$, $b_1=b_3=1/x$, $b_2=b_4=\lr{x-2}/x$, with $x=2-2^{1/3}$.\cite{Ruth_1983, McLachlan_1995, Grochowski_Lesyng_2003, BCM08, Lubich_2008} 
Note that $H_{\rm int}$ [Eq.\ (\ref{eq:hint})] only contains the interaction term with the oscillators and not the Hubbard-$U$, which ensures that no additional error will be introduced in the case $l=0$, which otherwise would result from splitting non-commuting terms of the Hamiltonian.

For our current problem, we have $T_{\text{el}} = T_{\text{el}}(t)$ due to the pump pulse, which must be seen as a complication. 
We choose a pragmatic solution by taking small time steps (typically, $\delta t=0.01$) for the time propagation during the pump pulse.
As this regards a short part of the propagation, it is not very expensive computationally.
Once the pulse is over, longer steps are chosen ($\delta t = 0.05$). 
As an error measure one can take the total energy, which becomes constant after the pump is over. 
The relative error $|E - \avg{H(t)}| / E$ is typically around $10^{-7}$ for the parameters above.

\section{Ground-state phase diagram of the dynamic Hubbard model}
\label{app:DynHubPD}

\begin{figure}[t]
\includegraphics[width=0.95\columnwidth]{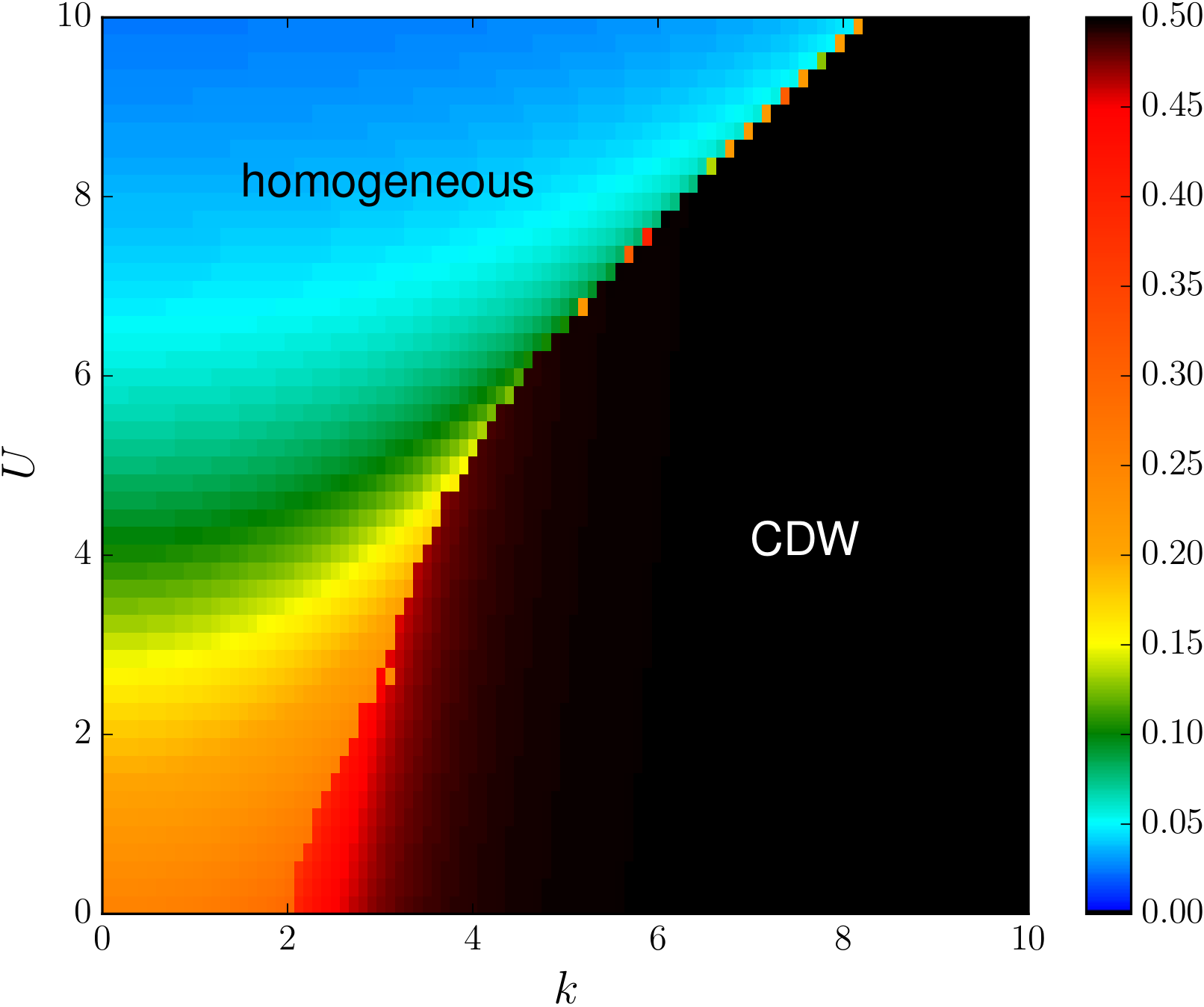}
\caption{
Equilibrium phase diagram of the dynamic Hubbard model.
The total double occupancy Eq.\ (\ref{eq:dtot}), see color code on the right, is displayed as a function of the electron-oscillator coupling strength $k$ and the Hubbard interaction $U$. 
Results for half-filling and for $L=10$ sites (periodic boundary conditions).  
In the CDW phase, the local double occupancy $\avg{n_{i\uparrow}n_{i\downarrow}}$ alternates between $0$ and $1$, leading to $d_{\text{tot}}=0.5$.
}
\label{fig:DynHubPD}
\end{figure}

In equilibrium, we have $\dot{p}_{i}=0$, and from Eq.\ (\ref{eq:motion_pq}) we can see that the displacement of the oscillators is given by $q_i^{(0)} = - k \avg{n_{i\uparrow}n_{i\downarrow}}$, i.e., the ground state has to be determined self-consistently. 
If the coupling strength $k$ becomes comparable to $U$, the system undergoes a phase transition to a symmetry-broken charge-density wave (CDW) phase, where every second site is doubly occupied. 
This maximizes the double occupancy and reduces the total energy compared to that of a homogeneous state. 
The $k$-$U$ phase diagram for a system of $L=10$ sites is shown in Fig.\ \ref{fig:DynHubPD}.
One observes that the phase boundary is very sharp despite the small system size.

\end{document}